\documentclass[aps,12pt,preprint,groupedaddress,showpacs,reprintnumbers,amsmath,amssymb,floatfix]{revtex4}
\usepackage{ulem} % added by FW \sout {...}
\usepackage{graphicx}
\usepackage{dcolumn}
\usepackage{bm}
\usepackage{epsf}
\usepackage{epsfig}
\usepackage{amsmath,amsfonts,amsthm, amssymb,latexsym}
\usepackage{enumerate}
\usepackage{color}

\newcommand{\beq}{\begin{equation}}
\newcommand{\eeq}{\end{equation}}
\newcommand{\bcn}{\begin{center}}
\newcommand{\ecn}{\end{center}}

\newcommand{\lsim}{\lower0.5ex\hbox{$\; \buildrel < \over \sim \;$}}

\begin{document}

\title{Braking Index of Isolated Pulsars II: A novel two-dipole model of pulsar magnetism.}

\date{\today}

\author{O.Hamil} \email{ohamil8@gmail.com}
\affiliation{Department of Physics and Astronomy, University of Tennessee, Knoxville, TN 37996, USA}
\author{N.J.Stone} \email{nick.stone@physics.ox.ac.uk}
\affiliation{Department of Physics and Astronomy, University of Tennessee, Knoxville, TN 37996, USA}
\affiliation{Department of Physics, Oxford University, Oxford UK}
\author{J.R.Stone} \email{jirina.stone@physics.ox.ac.uk}
\affiliation{Department of Physics and Astronomy, University of Tennessee, Knoxville, TN 37996, USA}
\affiliation{Department of Physics, Oxford University, Oxford UK}

\begin{abstract}
The magnetic dipole radiation (MDR) model is currently the best approach we have to explain pulsar radiation. However a most characteristic parameter of the observed radiation, the braking index n$_{\rm obs}$ shows deviations for all the eight best studied isolated pulsars, from the simple model prediction n$_{\rm dip}$ = 3. The index depends upon the rotational frequency and its first and second time derivatives, but also on the assumption of that the magnetic dipole moment and inclination angle, and the moment of inertia of the pulsar are constant in time. In a recent paper [Phys. Rev. D 91, 063007 (2015)] we showed conclusively that changes in the moment of inertia with frequency alone, cannot explain the observed braking indices.

 Possible observational evidence for the magnetic dipole moment migrating away from the rotational axis at a rate $\dot\alpha$ $\sim$ 0.6$^{\circ}$ per 100 years over the life time of the Crab pulsar has been recently suggested by Lyne et al. In this paper, we explore the MDR model with constant moment of inertia and magnetic dipole moment but variable inclination angle $\alpha$. We first discuss the effect of the variation of $\alpha$ on the observed braking indices and show they all can be understood. However, no explanation for the origin of the change in $\alpha$ is provided.

After discussion of the possible source(s) of magnetism in pulsars we propose a simple mechanism for the change in $\alpha$ based on a toy model in which the magnetic structure in pulsars consists of  two interacting dipoles. We show that such a system can explain the Crab observation and the measured braking indices. 

\end{abstract}


\maketitle

%\pacs{97.60.Jd, 97.60.Gb, 26.60.Dd,26.60.Gj, 26.60.Kp, 04.40.Dg, 04.25.D-}

\section{\label{sec1}Introduction}
There have been many attempts to explain the braking index extracted from observations of pulsar spin-down rates. To date, no accepted description exists for the rather wide ranging values (see Table~\ref{tab:1}) found for the eight  pulsars for which this parameter is reasonably well known. Most discussion starts with acknowledgement that there are three possible competing processes, mass emission (pulsar wind), magnetic dipole driven radiation and quadrupole radiation which, taken individually, would lead to braking indices $n$ of 1, 3 and 5, respectively \cite{ostriker1969}. It is generally agreed that the quadrupole radiation (n=5) may be neglected (all values of in Table~\ref{tab:1} are between 1 and 3) . However, combination of the other two mechanisms has been suggested \cite{alvarez2004}.  

Since the dominant feature of pulsar behavior is the magnetic dipole emission mechanism, it is logical to seek an alternative explanation of the observed braking indices based on the magnetic dipole description. The standard expression for the loss of energy due to magnetic dipole radiation is given in terms of the strength M of the dipole moment, the angle $\alpha$ between the pulsar rotational axis and the dipole axis, and the rotational frequency of the pulsar
\begin{equation}
\dot E = -{2\over3}M^2{\rm{sin^2}\alpha}\Omega^4.
\label{eq:00}
\end{equation}

Inserting rotational energy, we have

\begin{equation}
{d \over dt}\left({1\over2}{I\Omega^2}\right) = -{2\over3}M^2{\rm{sin^2}\alpha}\Omega^4,
\label{eq:01}
\end{equation}
and if MoI is assumed to be constant, we get the standard expression for time rate of change of frequency $\dot\Omega$, 

\begin{equation}
\dot\Omega\ = -\frac{2}{3}\frac{M^{\rm 2}{\rm sin^2\alpha}}{I}\Omega^{\rm 3}.
\label{eq:02}
\end{equation}
Using Equation~\ref{eq:01}, we obtain a general expression for  n$_{\rm obs}$ = $\ddot{\Omega} \Omega/\dot\Omega^{\rm 2}$, 
\begin{equation}
n_{\rm obs}=n_{\rm dip}+\frac{2\Omega}{\dot\Omega}\left(\frac{\dot\alpha}{{\rm tan}(\alpha)} + \frac{\dot M}{M}\right)
\label{eq:03}
\end{equation}
with n$_{\rm dip}$ = 3. 

In a previous paper \cite{hamil2015} some of us have explored the possibility that changes in the ellipticity of the pulsar caused by centrifugal stretching, and consequent changes in the MoI could contribute substantially. We found that, given the relatively slow rotational frequencies and the estimated properties of neutron star matter, such changes can produce no appreciable deviation from the MDR value n$_{\rm dip}$ = 3 for the known pulsars given in Table~\ref{tab:1}. 

Current understanding of pulsar magnetic fields does not allow for sound speculation as to a change in the strength of magnetic moment M. We are left with time variation of the inclination angle $\alpha$. Recent detailed analysis by Lyne et al. \cite{lyne2013,lyne2015} (LEA in future text) of the best and longest observation of any pulsar (the Crab) has revealed that the angle $\alpha$ for this star may be changing, albeit slowly, towards orthogonality, although LEA note that this is a model dependent interpretation of the observed data. Several authors (e.g. \cite{blandford1988,chen1998,allen1997,espinoza2011}) remarked on a possibility that a low braking index can be caused by an increasing the dipolar magnetic field or the change in the inclination angle.  Very recently,  Yi and Zhang \cite{Yi2015} showed that, in a model of the braking mechanism the time evolution of $\alpha$ could be of importance.

This paper is organized as follows. In Section~\ref{sec2} we explore implications of the MDR model with variable $\alpha$.  Magnetic properties of pulsars are briefly summarized in Section~\ref{sec3}, and the two-dipole toy model is introduced in  Section~\ref{sec4}, followed by discussion in Section~\ref{sec5}.

\section{\label{sec2}Braking index of a pulsar with changing inclination angle $\alpha$}
\label{sec:2}
In this section we explore the consequences of the $\dot\alpha$ dependent term in Equation~\ref{eq:03} before considering a possible mechanism. Table~\ref{tab:1} contains the relevant measured parameters n$_{\rm obs}$, the frequency $\Omega$ and its time derivative $\dot\Omega$. Rewriting Equation~\ref{eq:03} in the form
\begin{equation}
\frac{\dot\alpha}{{\rm tan}(\alpha)}=\frac{\dot\Omega}{\Omega}\left(\frac{n_{\rm obs}-n_{\rm dip}}{2}\right).
\label{eq:04}
\end{equation}
 yields values of the ratio $\dot\alpha$/tan($\alpha$), taking n$_{\rm dip}$ = 3.
Only for the Crab pulsar we have a value of the time variation of $\alpha$. Deducing n$_{\rm obs}$ from observation, LEA obtained  $\dot\alpha$ =  (0.566$\pm$0.002$)^{\rm 0}$/100 years taking $\alpha$ = (45$\pm$0.18)$^0$. This value is compatible with estimates of $\alpha$ being between 45$^{\rm 0}$ - 70$^{\rm 0}$  obtained from modeling the shape of the Crab beam (\cite{lyne2013} and refs therein). 

Neither the value of $\alpha$ nor that of $\dot\alpha$ is necessarily valid for other pulsars. However, to explore the possible range of these parameters, in Table~\ref{tab:1} we give the values of $\alpha$ found if all pulsars are taken to have the same $\dot\alpha$ as the Crab and, alternatively, the values of $\dot\alpha$ resulting from setting $\alpha$ to the Crab value.

\begin{figure}
\includegraphics[width=0.8\textwidth]{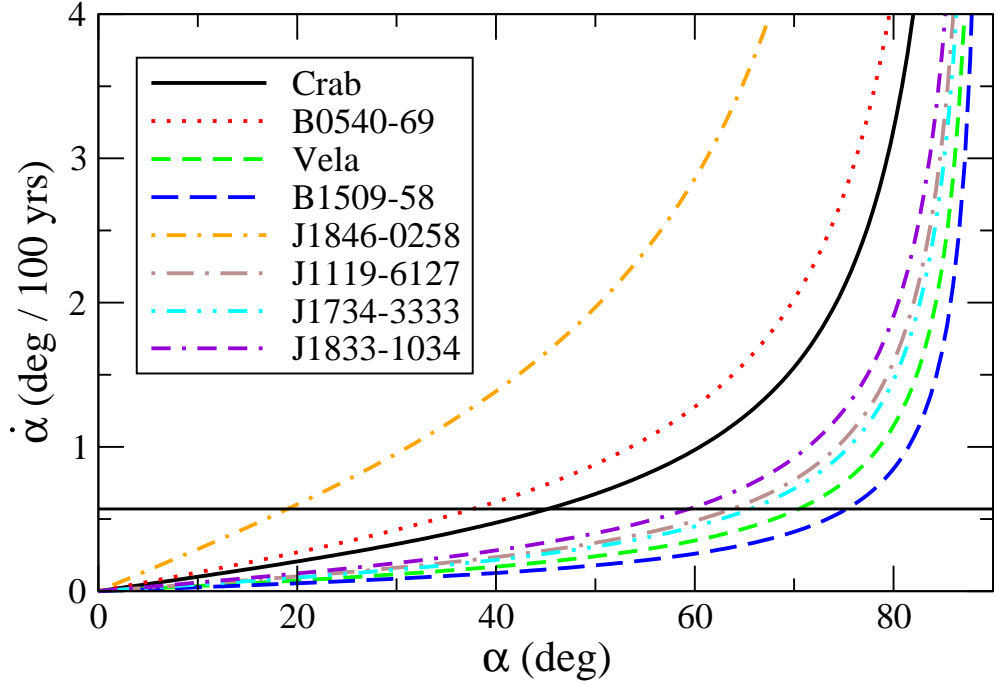}
\caption{\label{fig1}(Color on-line) Relation between the time rate of change in the inclination angle $\alpha$ and its magnitude, calculated using the measured braking index for the eight isolated pulsars in Table~\ref{tab:1}.  For more explanation see text.}
\end{figure}

The results show that all observed braking indices can be explained in terms of values of $\alpha$ and its time variation which are not very distant from those of the Crab, as illustrated in Fig.~\ref{fig1}. We show the relation between $\dot\alpha$ and $\alpha$ as calculated for eight well measured pulsars (including the Crab), taking data from Table~\ref{tab:1}.  The horizontal line shows the intersection of each curve with the measured value of $\dot\alpha$ from the Crab. It can be seen that for $\dot\alpha$ close the Crab value, all pulsars should have magnetic dipole oriented within 18 $ < \alpha <$ 80$^{\rm 0}$ of the axis of rotation. However, Fig.~\ref{fig1} shows only the case when $\alpha$, measured from the rotation axis to the north pole of the radiating dipole, lies in the first quadrant  between 0$^{\rm 0}$ - 90$^{\rm 0}$ where ${\rm tan}(\alpha)$  is positive.

In discussion of Equation~\ref{eq:04} and the values obtained in Table~\ref{tab:1}, we note that $\dot\Omega$ is always negative thus the sign of  $\dot\alpha$/tan($\alpha$) is determined by the sign of (n$_{\rm obs}$ - n$_{\rm dip}$). For all eight pulsars in Table~\ref{tab:1} this sign is negative and thus the sign of $\dot\alpha$/tan($\alpha$)  is positive. However, this does not determine the sign of either  $\dot\alpha$ or tan($\alpha$).

If $\alpha$ is taken as lying in either the first or third quadrant, (0$^{\rm 0}$ - 90$^{\rm 0}$ or 180$^{\rm 0}$ - 270$^{\rm 0}$) for which tan($\alpha$) is positive then $\dot\alpha$ is in turn positive. For $\alpha$ in either of these quadrants, positive $\dot\alpha$ means that the radiating dipole is rotating away from the pulsar rotation axis. However, if $\alpha$ lies in either of the second or forth quadrant ( 90$^0$ - 180$^0$ or 270$^0$ - 360$^0$), tan($\alpha$) is negative and thus $\dot\alpha$ is also negative. In this case the radiating dipole is moving towards the rotation axis. If a pulsar were found with (n$_{\rm obs} > $ 3), the sequence of possible tan($\alpha$) and $\dot\alpha$ signs would be inverted. We summarize all posibilites in Table~\ref{tab:0}. Note that the dipole emission power depends of sin$^{\rm 2}(\alpha)$ and is up/down symmetric, hence observations cannot reveal the orientation of the dipole. 
Examination of Table~\ref{tab:1} shows that braking indices of all eight pulsars can be explained by MDR alone provided the values of $\alpha$ are allowed to vary within 18$^{\rm 0} < \alpha <$ 80$^{\rm 0}$ and $\dot\alpha$ is taken as a constant in the time of observation.

\begin{table}
\centering
\caption{\label{tab:1} Observational data and calculated values of  $\dot\alpha/${\rm tan}($\alpha$) and $\alpha$ for eight best measured pulsars  \cite{lyne2015}.  The values of $\alpha$ required to fit the observed braking index taking $\dot\alpha$ = 0.56$^{\rm 0}$/100 years for all pulsars are given in column 6. The values of   $\dot\alpha/${\rm tan}($\alpha$) and $\alpha$ have the same percentage error as the quoted braking index. $\Omega$ and $\dot\Omega$ from \cite{lyne1993,livingstone2007,boyd1995,waltevrede2011,lyne1996,espinoza2011,roy2012}.}
\vspace{5pt}
\begin{tabular}{lcccccc}
\cline{1-6}
Pulsar                                              &          n$_{\rm obs}$                    &     $\Omega$     &  $\dot\Omega$                  & $\dot\alpha$/{\rm tan}($\alpha$)&     $\alpha$    \\ 
                                                        &                                     &      s$^{\rm -1}$    &  10$^{\rm -10} $s$^{\rm -2} $ &  per 100 years                           &    degrees      \\  \cline{1-6}
PSR B0531+21 (Crab)                       &  2.51$\pm$0.01          &  30.22543701     &  -3.862283                        &   0.566$\pm$0.002                &  45$\pm$0.18      \\
PSR B0540$-õµ$69           &  2.14$\pm$0.01        &  19.8344965       &  -1.88383                        &   0.738$\pm$0.003                     &  37$\pm$0.18        \\
PSR B0833$-õµ$45 (Vela) & 1.4$\pm$0.2               &  11.2                   &   -0.157                        &   0.20$\pm$0.03                          & 70$\pm$10       \\ 
PSR B1509$-$58                              &  2.839$\pm$0.001       &  6.633598804     &  -0.675801754              &   0.1482$\pm$0.0001                  & 75.32$\pm$0.03         \\   
PSR J1846$-õµ$0258        & 2.16$\pm$0.13           &  3.0621185502   &  -0.6664350                   &  1.6$\pm$0.1                           & 19$\pm$1         \\
PSR J1833$-õµ$1034        & 1.857$\pm$0.001   &  16.159357         & -0.5275017                        & 0.3371$\pm$0.0002                    &59.21$\pm$0.03   \\
PSR J1119$-$6127                           &  2.684$\pm$0.001      &  2.4512027814   & −0.2415507                  & 0.2814$\pm$0.0002                   &63.56$\pm$0.05           \\
PSR J1734$-$3333                           &  0.9$\pm$0.2              &  0.855182765     &  -0.0166702                  & 0.37$\pm$0.08                          & 57$\pm$12           \\    \cline{1-6}
\end{tabular}
\end{table}

Many authors correctly note that the existence of a co-rotating magnetospheric plasma should contribute to the overall energy loss of pulsars \cite{contopoulos2006, contopoulos2007, li2012}. This additional energy radiation  is a relativistic effect due to motion of charged particles in the magnetosphere, the particle wind. If acting alone, as would be the case for pulsars with the magnetic dipole aligned with the axis of rotation ($\alpha$ = 0), leads to n=1.

If both, the wind and pure magnetic dipole radiation contribute to the energy loss, a modified braking index between 1 and 3  can be calculated and could account for the observed braking indices \cite{alvarez2004}.  Our results show that the braking index over the entire range of observation, $0.9 < $n$_{\rm obs} < $2.8, can be explained by MDR in vacuum alone. This result may suggest that the effect of the particle wind is not significant to the first approximation, especially for values of $\alpha$ above about $20^{\rm 0}$.

\begin {table}
\centering
\caption{\label{tab:0} Sign of $\dot\alpha$/tan($\alpha$) and tan($\alpha$) combinations leading to the increase or decrease of $\dot\alpha$. The rotation axis is taken as taken at $\alpha$ = 0. Q stands for quadrant: [Q1] spans 0$^{\rm 0}$ - 90$^{\rm 0}$, [Q2] spans 90$^0$ - 180$^0$, [Q3] spans 180$^0$ - 270$^0$ and [Q4] spans  270$^0$ - 360$^0$ moving the north pole of the dipole clockwise from the rotational axis.  Combined positive (negative) sign means increase (decrease) in $\dot\alpha$ and a consequent movement away (towards) the axis of rotation. The left (right) part of the table is calculated using (n$_{\rm obs}$ - 3) negative (positive). }
\vspace{5pt}
\begin{tabular}{cccc|cccc}
\hline
 $\dot\alpha$/{\rm tan}($\alpha$) & {\rm tan}($\alpha$) &   Quadrant   &   $\dot\alpha$ &  $\dot\alpha$/{\rm tan}($\alpha$) & {\rm tan}($\alpha$) &   Quadrant   &   $\dot\alpha$ \\ \hline
     +          &    +     &    Q1     &    increase     &   --          &     +   &     Q1    &    decrease   \\
      +           &   +        &    Q3    &     increase  &      --     &     +   &     Q3    &    decrease  \\
      +         &     --    &     Q2     &    decrease      &      --     &      --   &     Q2      &   increase  \\
       +          &     --      &    Q4     &    decrease   &      --       &    --       &      Q4     &   increase \\ \hline
  \end{tabular}
\end {table}

\section{\label{sec3}Mechanisms of Pulsar Magnetism}
The origins and distributions of the magnetic fields of pulsars, and their misalignment with respect to the axis of rotation, are not well understood. There is extensive literature on this subject, documenting the complexity of the problem (see e.g. \cite{spruit2008, fujisawa2014, potekhin2014} ). 

Observational evidence for the intensity of magnetic fields in pulsars is also very limited. The only direct information comes from pulsars accreting material from a binary partner \cite{coburn2002} which have shown signals interpreted as cyclotron resonance involving electrons orbiting the field lines. The resonance frequencies correspond to fields  B $\sim$ 1.4 x 10$^{12}$ G \cite{reisenegger2003}. In isolated pulsars, the field is usually derived from the relation between the period of rotation P and its time rate of change $\dot{\rm{P}}$, assuming magnetic dipole radiation, using the formalism detailed in \cite{hamil2015}, which gives rise to Equation~\ref{eq:01}. The values obtained are also of the order of 10$^{12}$ G.  The existence of objects with extremely strong surface magnetic fields, up to 10$^{15-16}$ G, based on observation of high energy X-ray and gamma-rays, known as magnetars, seems to be generally accepted \cite{harding2006}.

The two principal potential sources of magnetic field in pulsars that are currently discussed are the dynamo effect, and constituent magnetization arising from the formation of ferromagnetically ordered matter. The dynamo theory describes the process through which a rotating, convecting, and electrically conducting fluid acts to maintain a magnetic field. It requires kinetic energy, which is provided by the pulsar rotation and an internal energy source to drive convective motions within the fluid \cite{thompson1993}. A dipole produced by this mechanism, essentially linked to the rotation of the pulsar, may be expected to be coaxial and centered within the star.  Existence of a stable, ferromagnetically ordered region inside the liquid interior of the pulsar has been  discussed by many authors in the past (see e.g. \cite{haensel1996} and ref. therein) and revived recently \cite{eto2013, hashimoto2015}.  With pulsar radii R of order 10$^{\rm 6}$ cm, the associated magnetic dipole moments (of order BR$^{\rm 3}$) are 10$^{\rm 30}$ ergs/G for ordinary pulsars and 10$^{\rm 33}$ ergs/G for magnetars.  Such huge magnetization could arise from the constituent nucleons, which number $\sim$ 10$^{\rm 57}$ in a typical pulsar of mass 1.5~M$_\odot$. Each has a moment of order 10$^{\rm -24}$ ergs/G, giving a potential total moment in broad agreement with the magnetar estimates \cite{harding2006}. However, the ordered material, mainly located in the core, may have a domain structure which could lead to reduced, local dipoles, neither coaxial nor con-centric with the rotation of the star.

Although the magnetic fields may have complicated intrinsic configurations including poloidal and toroidal components \cite{fujisawa2014}, they are likely dominated by a dipolar term to first approximation.  The assumption of a dipole explains the observed pulse, and the estimated power radiated due to rotation is correct in order of magnitude. However, there is the possibility that the observed dipole radiation is the resultant of more than one dipole. For example, coexistence of a dipole field due to the dynamo effect, and a field created by the spin-alignment of particles leading to formation of ordered domains could be a possible representation of such a configuration. Other sources of dipoles, such as motion of charged particles in the magnetosphere \cite{michel1991} cannot be excluded, but given the much lower density of the magnetosphere, the resulting moments and fields are likely to be small compared to those in the star. In this paper we consider possible situations involving two dipoles in the pulsar, generated by  different mechanisms.

\section{\label{sec4}Toy model of two interacting dipoles}
In Sec.~\ref{sec:2} we showed that the MDR model allows for increase or decrease of $\alpha$ in dependence of the sign of tan($\alpha$). However, the model does not offer any mechanism causing these changes.
Here we present a toy model which makes a crude first order attempt to understand the possible physics behind the change in $\alpha$ over time, as seen by LEA, which, in turn, can account for the braking index of the Crab pulsar, and the other observed pulsars in Table~\ref{tab:1}.

The model consists of two dipoles with magnetic moments $\vec{M_{\rm 1}}$ and $\vec{M_{\rm 2}}$, separated in space by distance $r$ with constant magnitude. Whilst the more familiar result of dipole - dipole interactions is an attractive ($\grave{a}$ la Van der Waals) or repulsive force (depending on their relative orientation) in the present context we focus on the potential effect of the turning moment, or couple, they exert on each other.  We hold their separation constant, thus neglecting the effect of the linear force between the dipoles, and assume that one dipole ($\vec{M_{\rm 1}}$) is fixed at the center of the star and aligned with the rotation axis, consistent with the dynamo mechanism. The line joining the centers of the two dipoles (the dipole-dipole axis) makes an angle $\theta_{\rm 1}$  with the rotation axis. We further assume that $\vec{M_{\rm 2}}$ is initially coplanar with $\vec{M_{\rm 1}}$, thus eliminating all azimuthal angles $\phi$ from the problem.  The second dipole $\vec{M_{\rm 2}}$ is free to rotate about its center with angle $\theta_{\rm 2}$ measured from the dipole-dipole axis to $\vec{M_{\rm 2}}$.

The magnetic field generated by dipole $\vec{M_{\rm 1}}$ at position $r$ is
\begin{eqnarray}
\vec{B_{\rm 1}} = {\mu_0 \over {4\pi}}{1 \over r^{\rm 3}}[3(\vec{M_{\rm 1}}\cdot \hat{r})\hat{r} - \vec{M_{\rm 1}}],
\label{eq:05}
\end{eqnarray}
where $\mu_{\rm 0}$ is the magnetic permeability of free space and $\hat{r}$ is a unit vector along $r$. The potential energy of a dipole $\vec{M_{\rm 2}}$, in the magnetic field  $\vec{B_{\rm 1}}$ is given by
\begin{eqnarray}
U_{\rm {21}} = - \vec{M_{\rm 2} }\cdot \vec{B_{\rm 1}},
\label{eq:06}
\end{eqnarray}
that is
\begin{eqnarray}
U = -{\mu_0 \over {4\pi}}{1 \over r^{\rm 3}}[3(\vec{M_{\rm _1}}\cdot\hat{r})(\vec{M_{\rm_2}}\cdot\hat{r})- \vec{M_{\rm 1}}\cdot\vec{M_{\rm 2}}].
\label{eq:07}
\end{eqnarray}
With the limitation mentioned above, the potential reduces to
\begin{eqnarray}
U( \theta_{\rm 1}, \theta_{\rm 2})= {{\mu_{\rm 0} M_{\rm 1} M_{\rm 2}} \over {4\pi r^{\rm 3}}}(\sin{\theta_{\rm 1}}\sin{\theta_{\rm 2}} - 2\cos{\theta_{\rm 1}}\cos{\theta_{\rm 2}}).
\label{eq:08}
\end{eqnarray}
with $r$ held constant.
In the following sections, we examine the variation of U($\theta_{\rm 1}, \theta_{\rm 2}$) for different initial values of $\theta_{\rm 1}$ and $\theta_{\rm 2}$ to explore possible motion of $\vec{M_{\rm 2}}$, followed by a calculation of the couple acting on  $\vec{M_{\rm 2}}$, and possible dynamics of the resulting motion.

\begin{figure}
\includegraphics[width=0.7\textwidth]{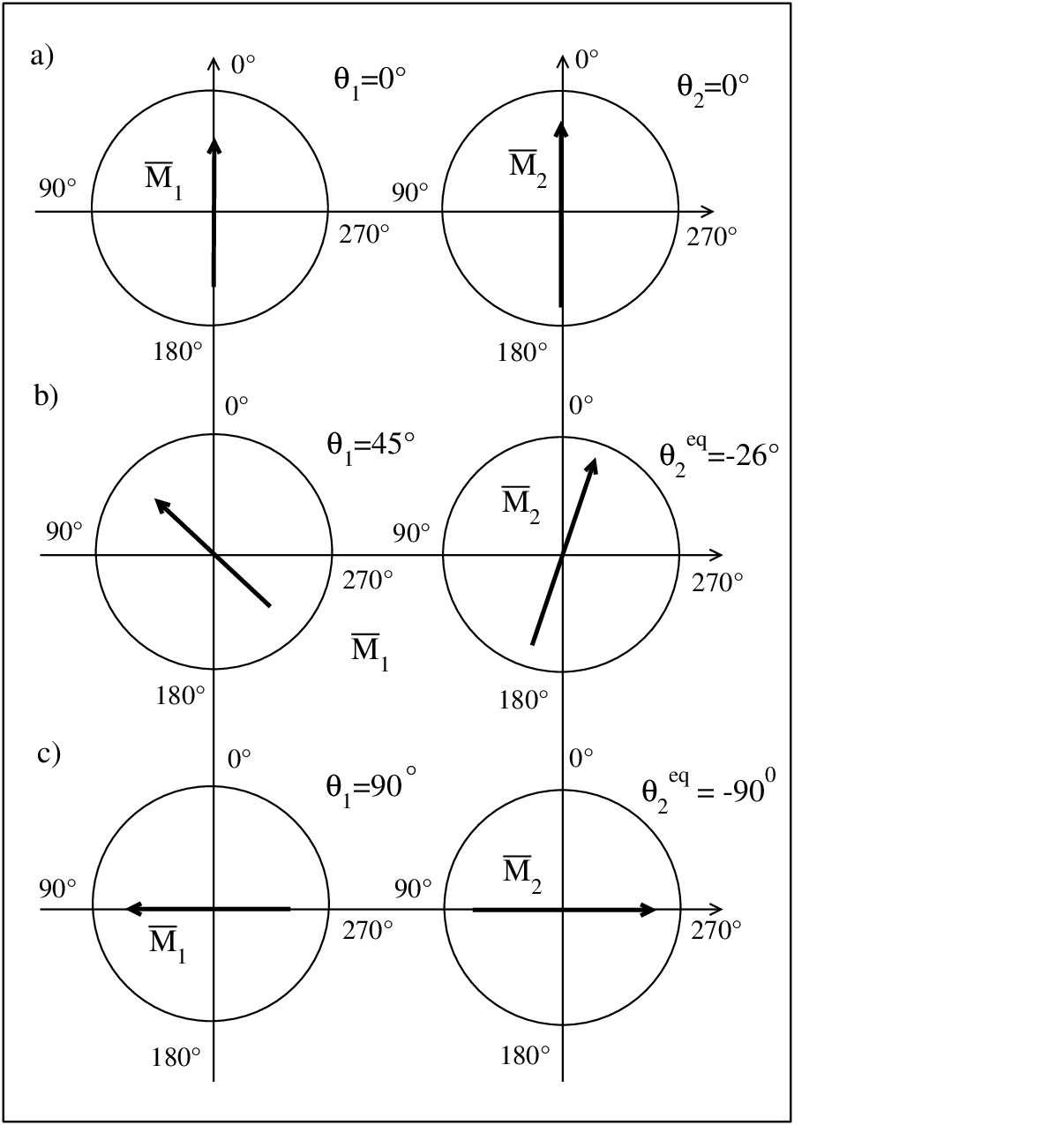}
\caption{\label{fig2} (a -c) Examples of configurations of two coplanar dipoles with fixed magnitude and distance between their origin in free space (a - c). The positive (negative) angles $\theta_{\rm 1}$ and $\theta_{\rm 2}$ are measured from the dipole-dipole axis anti-clockwise (clockwise).}
\end{figure}

\begin{figure}
\includegraphics[width=0.7\textwidth]{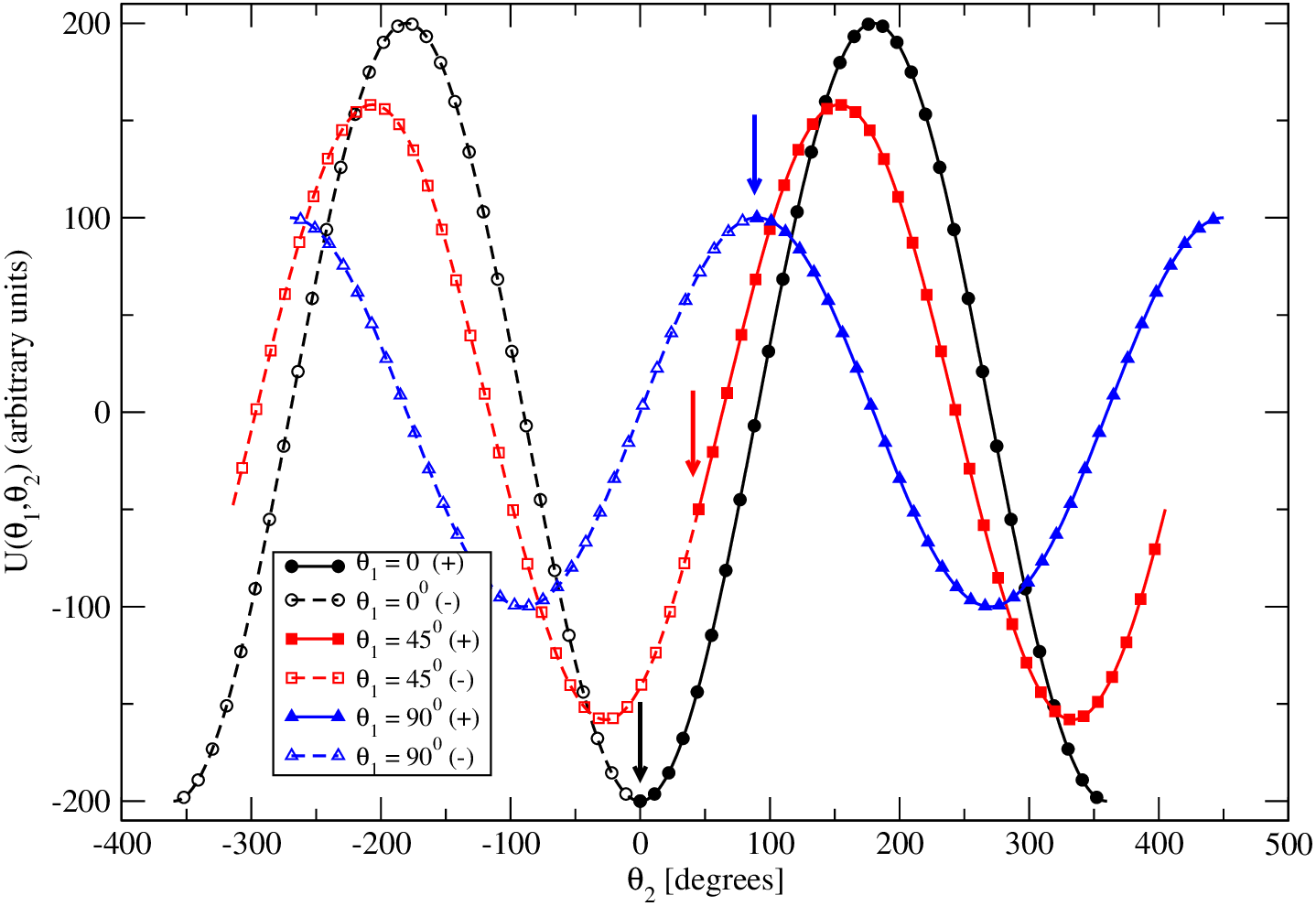}
\caption{\label{fig3}(Color on-line) Potential energy U( $\theta_{\rm 1}, \theta_{\rm 2}$) as a function of $\theta_{\rm 2}$ governing the development of the two-dipole systems for three sets of  initial conditions $\theta_{\rm 1}$ = $\theta_{\rm 2}$ = 0$^{\rm 0}$, 45$^{\rm 0}$ and 90$^{\rm 0}$. The arrows indicate initial values of $\theta_{\rm 2}$. The +(-) signs in the legend indicate anti-clockwise (clockwise) change in $\theta_{\rm 2}$.  Friction in the material of the pulsar is neglected. For more discussion see text.}
\end{figure}

In Fig.~\ref{fig2} a-c, three different initial alignments of the two dipole are taken to illustrate possible examples of the variation of U( $\theta_{\rm 1}, \theta_{\rm 2}$) with  $\theta_{\rm 2}$ for different  $\theta_{\rm 1}$. The variation of U( $\theta_{\rm 1}, \theta_{\rm 2}$)  as a function of  $\theta_{\rm 2}$ is shown in Fig.~\ref{fig3} for each example. All have the same sinusoidal oscillatory behavior with a single, stable, minimum energy value of  $\theta_{\rm 2}$. Clearly, for an arbitrary starting point, $\vec{M_{\rm_2}}$ will rotate towards this minimum state. For small initial displacements from the minimum, the motion is simple harmonic, but it is more complex, although oscillatory, for other starting points. The direction of rotation depends upon the initial value of the difference between $\theta_{\rm 1}$ and $\theta_{\rm 2}$.

\begin{figure}
\includegraphics[width=0.7\textwidth]{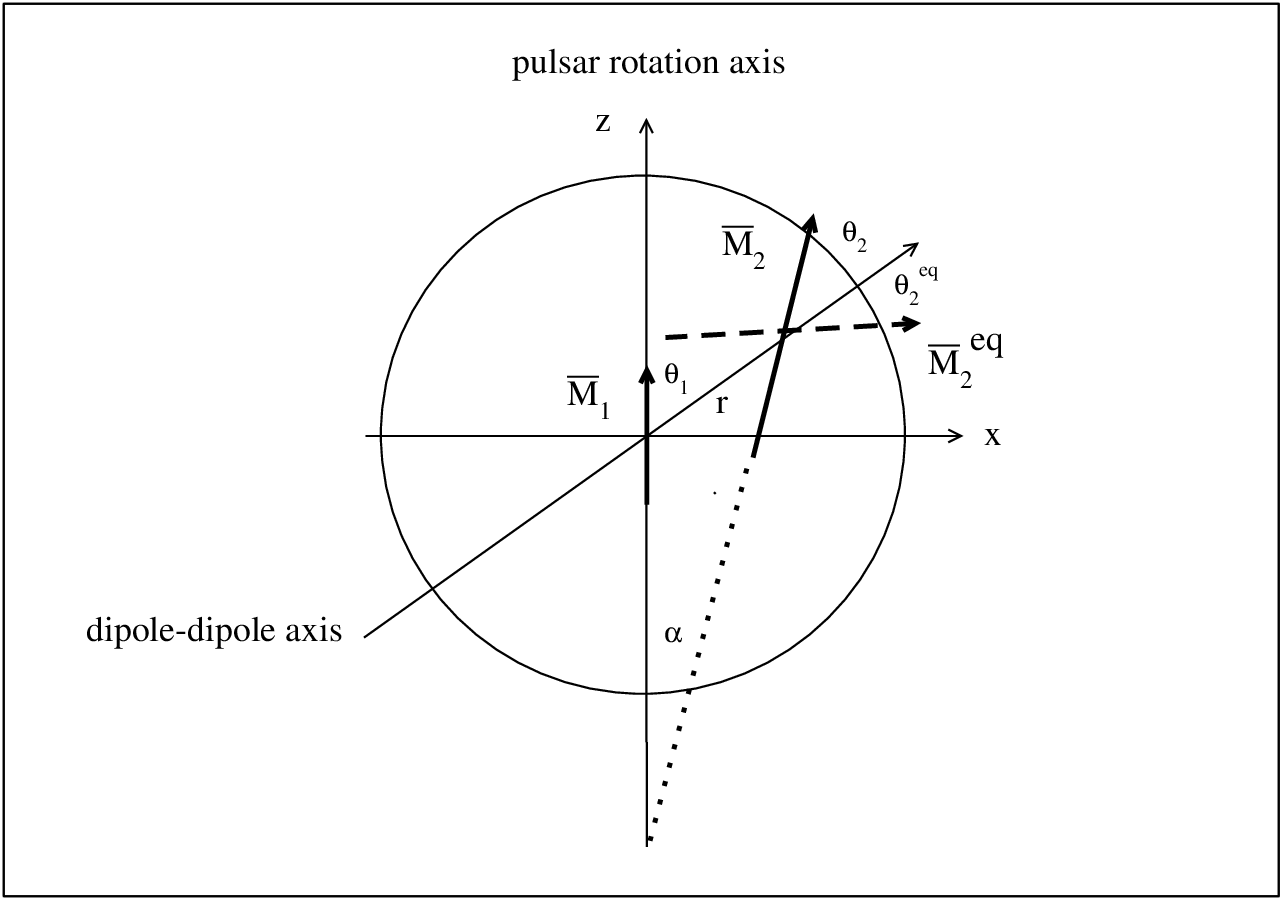}
\caption{\label{fig4} The two-dipole system in the pulsar. The frame of reference has origin in the center of the star, where $\vec{M_{\rm 1}}$ is fixed and aligned with the rotations axis $z$. The dipole-dipole axis $r$ makes an angle  $\theta_{\rm 1}$ with $z$. $\alpha$ is an angle between $z$ and $\vec{M_{\rm 2}}$ measured from $\vec{M_{\rm 2}}$. The dashed line corresponds to $\vec{M_{\rm 2}}$ in the equilibrium position. For more explanation see text}.
\end{figure}

In Fig.~\ref{fig4}, we show the general arrangement of $\vec{M_{\rm 1}}$ and $\vec{M_{\rm 2}}$ in a pulsar with $\vec{M_{\rm 1}}$ along the axis of rotation of the star, and the dipole-dipole axis at angle $\theta_{\rm 1}$ to the axis of rotation. $\vec{M_{\rm 2}}$ is set at angle  $\theta_{\rm 2}$ to the dipole-dipole axis, so that, with the angle $\alpha$ as defined in the pulsar literature, we have $\alpha$ =  $\theta_{\rm 1}$ -  $\theta_{\rm 2}$ and $\dot\alpha$ = -$\dot\theta_{\rm 2}$.

The actual motion of $\vec{M_{\rm 2}}$ will depend upon the magnitude of the couple acting, the MoI of the rotating region, and the resistance to motion.  An expression for the couple C 
\begin{equation}
C = -\frac{\partial U}{\partial \theta_{\rm 2}} =  {{\mu_{\rm 0} M_{\rm 1} M_{\rm 2}} \over {4\pi r^{\rm 3}}}({\rm sin}{\theta_{\rm 1}}{\rm cos}{\theta_{\rm 2}}+2 {\rm cos}{\theta_{\rm 1}}{\rm sin}{\theta_{\rm 2}})= I_{\rm 2}\ddot\theta_{\rm 2}
\end{equation}
is a somewhat complicated, but generaly non-zero, function of $\theta_{\rm 1}$ and $\theta_{\rm 2}$.
We briefly present two alternative versions of the toy model which may exist, given the limitations of our understanding of neutron star interiors, including the possibility of superfluidity.  The first version omits friction, hence allowing accelerated rotation. The kinetic energy acquired by the region of the star supporting $\vec{M_{\rm 2}}$ is then

\begin{eqnarray}
{1 \over 2} I_{\rm 2} \dot\theta_{\rm 2}^2 = \Delta U = U_{\rm i} - U_{\rm f}.
\label{eq:09}
\end{eqnarray}
where $i$, and $f$ denote the initial and final values of $\theta_{\rm 2}$ respectively.

Using (\ref{eq:08}) with re-arrangement and expressing the results in $cgs$ units, setting $\mu_{\rm 0}$/4$\pi$ equal to one, we can write
\begin{eqnarray}
\dot\theta_{\rm 2}^2 = {{2 M_{\rm 1} M_{\rm 2}} \over {I_{\rm 2} r^{\rm 3}}} F(\Theta).
\label{eq:010}
\end{eqnarray}
where
\begin{eqnarray}
F(\Theta) =(\sin{\theta_{\rm 1i}}\sin{\theta_{\rm 2i}} - 2\cos{\theta_{\rm 1i}}\cos{\theta_{\rm 2i}}) - (\sin{\theta_{\rm 1f}}\sin{\theta_{\rm 2f}} - 2\cos{\theta_{\rm 1f}}\cos{\theta_{\rm 2f}})
\label{eq:011}
\end{eqnarray}
is a general function, introduced for convenience, with $\Theta$ standing for all angles appearing in (\ref{eq:08}) and I$_{\rm 2}$ is the moment of inertia of the rotating material.

To estimate expected values of $\vec{M_{\rm 1}}$ requires, as input, the value of $\theta_{\rm 1i}$ (=$\theta_{\rm 1f }$) and the initial and final values of $\theta_{\rm 2}$.  We use the only example for which we have data, the Crab pulsar. Since the total change in $\theta_{\rm 2}$ turns out to be small, we can assume approximately constant angular acceleration so the average angular velocity is half its final (current) value which is $\sim$0.6$^{\rm o}$/100 years. Thus over the full lifetime of the Crab pulsar, $\theta_{\rm 2}$ will have changed by approximately 3$^{\rm o}$  from $\theta_{\rm 2i}$ to its current value $\theta_{\rm 2f}$.  If we choose  $\theta_{\rm 1i}$ =  45$^{\rm o}$ then $\theta_{\rm 2i}$ = 3$^{\rm o}$ and $\theta_{\rm 2f}$ = 0$^{\rm o}$ (aligned with the dipole-dipole axis).

These $\theta$ values give an estimated value of F($\Theta$) $\sim$ 0.0389. Taking B $\sim$ 10$^{\rm 12}$ G, R $\sim$ 10$^{\rm 6}$ cm and r/R at a mid-value $\sim$ 0.5,  we can estimate the magnitude of the relevant parameter, ${M_{\rm 1}}$/I$_{\rm 2}$,  to be $\sim$10$^{\rm -35}$ G$^{\rm -1}$s$^{\rm -2}$ (see Equation~\ref{eq:010}). The MoI of a slowly rotating neutron star is of order 10$^{\rm 46}$ g cm$^{\rm 2}$ \cite{hamil2015}.  Even taking I$_{\rm 2}$ as large as 10$^{\rm 44}$ g cm$^{\rm 2}$ (1 percent of the total MoI) would give, for  the magnitude of $\vec{M_{\rm 1}}$,  the relatively  small value $\sim$10$^{\rm 9}$ erg/G.

In this analysis, we have assumed that, for the Crab pulsar, the radiating dipole $\vec{M_{\rm 2}}$ is still in the initial phase of its motion and has not passed its minimum energy position. We cannot make the same assumption for the other pulsars, as their ages are unknown.  Furthermore, we cannot make any assumption about the total motion of $\theta_{\rm2}$ over the unknown lifetime of any pulsar other than the Crab, and therefore limit our order of magnitude calculations to the perceived motion on the order of time of the braking index measurements (i.e. 100 years) as an estimate of F($\Theta$).  

Taking the $\dot\alpha$ values from Table~\ref{tab:1} as indicating their present motion, and assuming these values to be relatively constant over 100 years, we can estimate the range of values of $\vec{M_{\rm 1}}$ for $\alpha$ = 45$^{\rm o}$ and differing values of r/R.  These values span from roughly (10$^{\rm8}$ -- 10$^{\rm12}$) erg/G for the remaining pulsars.  All values have considerable uncertainty resulting from the order of magnitude assumptions made, but are similar in magnitude to the estimation of the Crab pulsar.  In comparison with the maximum known pulsar magnetization, $\sim$10$^{\rm 33}$ erg/G, all results for the magnitude of $\vec{M_{\rm 1}}$ are very small.  The magnitude of $\vec{M_{\rm 1}}$, for any single pulsar, cannot be made more than two orders larger by variation of $r$, and is still many orders smaller than the value of $\vec{M_{\rm 2}}$ taken to be the source of the observed radiation.

It may be difficult to accept the idea of such a small driving couple, but recall that the model neglects friction, and that the angular velocity, even in so massive a body as the pulsar, is only $\sim$ 0.6$^{\rm 0}$ per century, which corresponds to a rotational period of $\sim$ 60,000 years. We therefore introduce a second version of the model in which we consider the motion to be friction limited. The action of friction, as it affects motion, can be represented in many ways.  For example, the convention in damped simple harmonic motion of considering  a resistance proportional to system velocity where, as for a block on an inclined plane, static friction proportional to the normal reaction may, or may not, be sufficient to prevent all motion.

To make this point clear we take, as an example, a constant frictional force. Conservation of energy from the start of the motion of $\vec{M_{\rm 2}}$ then gives 
\begin{equation}
{1 \over 2} I_{\rm 2} \dot\theta_{\rm 2}^2 + K(\theta_{\rm 2i} - \theta_{\rm 2f}) = U_{\rm i} - U_{\rm f} =  \Delta  U .
\end{equation}
where $K$ is the energy loss per unit turn angle. The equation for $\dot\theta_{\rm 2}^2$ becomes
\begin{equation}
\dot\theta_{\rm 2}^2 =\frac{2}{I_{\rm 2}}\left(\frac{ M_{\rm 1} M_{\rm 2}}{ r^{\rm 3}} F(\Theta)-K\Delta\theta_{\rm 2}\right).
\end{equation}
and for the acceleration $\ddot\theta_{\rm 2}^2$
\begin{equation}
\ddot\theta_{\rm 2}=\frac{1}{I_{\rm 2}}\left(\frac{ M_{\rm 1} M_{\rm 2}}{r^{\rm 3}} \frac{\partial{F(\Theta)}}{\partial\theta_{\rm 2}}-K\right).
\label{3.1}
\end{equation}
so that in this scenario, it is clear that the acceleration is reduced and becomes zero when K = $\partial{U_{\rm f}}/\partial{\theta_{\rm 2}}$.  Since $\partial{U_{\rm f}}/\partial{\theta_{\rm 2}}$ is a function of $\theta_{\rm 2}$  the motion will reach an asymptotic steady velocity. If we assume that  $\vec{M_{\rm 2}}$ in the Crab pulsar has reached a friction limited angular velocity after only $\sim$1000 years, we may believe the same is true of other pulsars, and that all their angular velocities are in the same range as that observed for the Crab pulsar - about 1$^{\rm o}$ per century. In this version of the model, the magnitude of $\vec{M_{\rm 1}}$ is
\begin{equation}
M_1 = \left( {{{1}\over{2}}I_2\dot\theta_2^2}+{K\Delta\theta_2} \right){{r^3}\over{M_2 F(\Theta)}}
 %M_{\rm 1} =\left(\frac{I_{\rm 2}}{2}\dot\theta_{\rm2}^2 + K\Delta\theta_{\rm 2}\right)\frac{ r^3}{M_{2}F(\Theta)}.
\end{equation}
We know of no way to estimate $K$ in such an unknown medium as pulsar material, but it is clear that, in this scenario, the motion would be slower, and its initiation would require larger values $\vec{M_{\rm 1}}$. There are many other possible scenarios including those in which the motion does not start, or may start and subsequently come to rest.  However, the friction limited model allows for constant $\dot\alpha$ as observed over the $\sim$ 40 years of observation. The MDR model is then able to account for all of the reliably known braking indices, as discussed in Section~\ref{sec2}.  

As a further comment, noting again that $\vec{M_{\rm 2}}$ for pulsars is only a fraction (of order of 10$^{\rm -3}$) of the magnetization of magnetars, our concept is that $\vec{M_{\rm 2}}$ is a result of possible domain structure in the star core and does not include effect of the crust. The rotating (small volume) magnetized medium may be superfluid neutron pairs. Once again we are not aware of any mechanism to estimate rotational friction in such a system.

\section{\label{sec5}Discussion}\label{disc}
The origins of magnetism in stars and planets in general, and of pulsars in particular, is a widely discussed topic. While the orders of magnitude of estimated magnetar and pulsar moments can be understood as reasonable compared with the potential moments produced by ordered alignment of the stellar constituents, possibly augmented by the dynamo effect of circulating conducting charged material, no detailed understanding has yet emerged.
LEA recently reported that available data on the pulse structure of the crab pulsar suggests the angle of inclination between the dipole moment and rotational axis of the star is increasing. In the light of this finding, we propose that the mechanism of magnetism in pulsars involves two interacting dipoles, and their interaction produces the observed rotation shown by LEA. The physics of their production may be either or both of the dynamo effect and ordering of the intrinsic magnetic moments of the constituents of the star.  An attractive possibility is to associate one dipole with the dynamo effect which is co-rotating with the star and located at its center, with a second dipole, caused by intrinsic alignment and being the source of emitted radiation from the star, at a non-central, off-axis position.  A simple toy model is presented in two versions, both with the center dipole pinned to the axis, one in which the second, off center, dipole can rotate without friction, and the second in which the motion is friction limited.  It is shown that this single mechanism can explain the braking index of all eight well observed pulsars if a relatively slow variation of the angle $\alpha$ between the axis of the radiating dipole and the rotational axis of the star is accepted.  Basing our estimates on the LEA interpretation of observations of the Crab pulsar, we have shown, using the toy model with approximations, that the change in the angle $\alpha$ can be reproduced.  We have shown further that, if friction is neglected (as might follow in a superfluid scenario), the central dipole required to produce the observed variation of $\alpha$,  has a very small value as compared with, for example, full alignment of the constituents of the star.  Making the variation of $\alpha$ friction limited will lead to an increased central moment requirement, but without knowledge of the friction mechanism we cannot estimate how large this would become.  If the motion were without friction it should be periodic about some equilibrium orientation of the two dipoles. Friction changes this picture.  It may take centuries of observation to establish a complete picture, which may differ between pulsars.  While many details are missing, we consider the possible two-dipole mechanism suggested here as offering a  significant outline explanation of the more salient facts concerning pulsar magnetism. 

We note that the toy model is, strictly speaking, applicable for an isolated pulsar in vacuum. The effect of co-rotating plasma filling the magnetosphere on the spin-down of pulsars has been studied since late 1960’s (see e.g. \cite{goldreich1969} and more recently \cite{contopoulos2006,contopoulos2007,li2012}). The main consequence of this effect is that it causes additional dissipation of energy from pulsars even when the magnetic axis is aligned with the rotation axis, and thus emission of magnetic dipole radiation does not occur. However, as stated \cite{contopoulos2006}, the exact theory of  this phenomenon is not yet available, and only empirical approximations were explored.

In a system of two dipoles there may be a small quadrupole contribution. The resultant of such an arrangement is a small change in the dominant dipole direction but little change in its magnitude, combined with a small quadrupole moment set perpendicular to the resultant dipole axis. The power radiated by such a quadrupole, being composed of two opposing dipoles, must be negligible compared with with the dominant dipole radiation and as such would not affect our results.

We have shown that the toy model allows, in principle, both increasing and decreasing $\alpha$, from and towards the rotation axis over time, and can explain braking indices both lower and higher than the canonical value of three yielded by the static MDR model. This feature may be interesting to follow in the light of recent observation of the braking index of PSRJ1640–4631 n$_{\rm obs}$ = 3.15$\pm$0.03 reported by Archibald et al. \cite{archibald2016} if it is confirmed.

\bigskip
%\vspace{15cm}

{\bf Acknowledgement.}
OH gratefully acknowledges partial support by the Department of Physics and Astronomy of the University of Tennessee in the course of this work.

\end{document}